----------------------------- Start of body part 2

\documentstyle[12pt,epsf]{article}

\begin{document}

\begin{titlepage}

\title{Quantum Clifford-Hopf Algebras for Even Dimensions}

\author{E. L\'opez \\
{\it Instituto de Matem\'aticas y F\'{\i}sica Fundamental, CSIC} \\
{\it Serrano 123, E--28006 Madrid, SPAIN}}

\date{}

\maketitle

\begin{abstract}
In this paper we study the quantum Clifford-Hopf algebras
 $\widehat{CH_q(D)}$
for even dimensions $D$ and obtain their intertwiner
$R-$matrices, which are
elliptic solutions to the Yang-Baxter equation. In the
trigonometric limit of
these new algebras we find the possibility to connect with
extended supersymmetry. We also analyze the corresponding spin
chain hamiltonian,
which leads to Suzuki's generalized $XY$ model.
\end{abstract}

\vspace{5cm}

\end{titlepage}

\subsection*{1. Introduction}
\indent

The quantum group structure plays an important role in the study
of two dimensional
integrable models because $R$-matrices intertwining between
diferent
irreps of a quantum group provide solutions to the Yang-Baxter
equation.
Two important families of integrable models are the 6-vertex and
8-vertex solutions to the Yang-Baxter equation \cite{bax}.
Whereas the 6-vertex
solutions are intertwiners $R$-matrices for
$U_q(\widehat{sl(2)})$,
a quantum group interpretation for the elliptic 8-vertex
family is not yet known.

Nevertheless, the 8-vertex regime is well understood for the
particular class of solutions to the Yang-Baxter equation
satisfying the free-fermion condition \cite{FW}
\begin{equation}
R_{00}^{00}(u) R_{11}^{11}(u) + R_{01}^{10}(u) R_{10}^{01}(u) =
R_{00}^{11}(u) R_{00}^{11}(u) + R_{01}^{01}(u) R_{10}^{10}(u)
\end{equation}
Indeed, a quantum group like structure has been found recently
for the most general free
fermionic elliptic 8-vertex model in a magnetic field.
The matrix of its Boltzmann weights \cite{F,BS}
acts as intertwiner for the afinization of a quantum Hopf
deformation of the Clifford algebra in two dimensions, noted
$\widehat{CH_q(2)}$ \cite{U}.

A major interest of the free fermionic solutions to the
Yang-Baxter equation is in their connection, in the 6-vertex
limit ($R_{00}^{11}(u)=R_{11}^{00}=0$),
with $N=2$ supersymmetric
integrable models.
The free fermionic 6-vertex solutions
are given by the $R-$matrix intertwiners between nilpotent
irreps
of the Hopf algebra $U_{\epsilon}(\widehat{sl(2)})$, with
$\epsilon^4=1$
(the nilpotent irreps are a special case of the cyclic
representations that enlarge the representation theory of
$U_{\epsilon}(\widehat{sl(2)})$
when $\epsilon$ is a root of unity).
In the trigonometric limit
the $R-$matrix for $\widehat{CH_q(2)}$ becomes
that for $U_{\epsilon}(\widehat{sl(2)})$, $\epsilon^4=1$.

In this article we construct the quantum Clifford-Hopf
algebras $\widehat{CH_q(D)}$ for even dimensions $D \geq 2$,
generalizing the results in \cite{U}. This general case is
interesting because it yields one of the rare examples of
elliptic $R-$matrices. The $R-$matrices we find admit
several spectral parameters, due to the structure of
$\widehat{CH_q(D)}$ as a Drinfeld twist \cite{D} of the
tensor product of
several copies of $\widehat{CH_q(2)}$.
The possibility to connect with
extended supersymmetry in the trigonometric limit of
$\widehat{CH_q(D)}$,
and a related supersymmetric integrable model
are analyzed in sect.3. Finally, in sect.4,
we study the spin chain hamiltonian associated to these
algebras. The model obtained represents
several $XY$ Heisenberg chains in an external magnetic field
\cite{LSM}
coupled among them in a simple way. Though the
coupling is simple it can be an starting point to get a
quantum group structure for more complicated models built
through the coupling of two XY or XX models
(Bariev model \cite{B}, 1-dimensional Hubbard model).
The last part of this section is devoted
to showing the equivalence  of this model --under some
restrictions-- with
a generalized $XY$ model proposed by M.Suzuki
in relation with the 2-dimensional dimer problem \cite{S}.

\subsection*{2. The quantum Clifford algebra}
\indent

A Clifford algebra $C(\eta)$ related to a cuadratic form or
metric $\eta$ is the associative algebra generated by the
elements $\{
\Gamma_{\mu} \}_{\mu =1}^D$, which satisfy
\begin{equation}
\{ \Gamma_{\mu}, \Gamma_{\nu} \} = 2 \eta_{\mu \nu} {\bf 1}
\;\;\; \mu, \nu = 1,\ldots,D
\label{8}
\end{equation}
\indent
The quantum Clifford-Hopf algebra $CH_q(D)$ \cite{U} is a
generalization and quantum deformation of $C(\eta)$,
generated by elements
$\Gamma_\mu$, $\Gamma_{D+1}$ (the analog of $\gamma_{5}$ for
the Dirac matrices) and new central
elements $E_\mu$ ($\mu =1,..,D$) verifying
\begin{eqnarray}
& & \Gamma_{\mu}^2 = \frac{q^{E_{\mu}}-q^{-E_{\mu}}}
{q- q^{-1}} \;\; , \;\; \Gamma_{D+1}^2 = {\bf 1}
\nonumber \\
& & \{ \Gamma_{\mu}, \Gamma_{\nu} \} =0, \;\; \mu \neq \nu
\nonumber \\
& & \{\Gamma_{\mu}, \Gamma_{D+1} \} =0 \label{9} \\
& & [ E_{\mu}, \Gamma_{\nu} ] = [ E_{\mu}, \Gamma_{D+1} ] =
[ E_{\mu}, E_{\nu} ] = 0 \;\; \forall \mu, \nu \nonumber
\end{eqnarray}
The charges $E_{\mu}$ result from elevating the components
of the metric $\eta$ from numbers to operators. The
generator $\Gamma_{D+1}$ will plays a similar role to
$(-1)^{F}$, with $F$ the fermion number operator. Although
for the standard Clifford algebra $D$ represents the
dimension of the space-time, in our case $D$ is only a
parameter labeling (3).
The algebra $CH_q(D)$ is a Hopf algebra
with the following comultiplication $\Delta$, antipode $S$
and counit $\epsilon$
\begin{equation}
\begin{array}{lll}
\Delta (E_{\mu}) = E_{\mu} \otimes {\bf 1} \: + \: {\bf 1}
\otimes E_{\mu}, \: \: \:& S(E_{\mu}) = -E_{\mu}, \: \:
&\epsilon(E_{\mu}) = 0 \\
\Delta (\Gamma_{\mu}) = q^{E_{\mu}/2} \Gamma_{D+1} \otimes
\Gamma_{\mu} \: +
\: \Gamma_{\mu} \otimes q^{-E_{\mu}/2}, \: \: \:
& S(\Gamma_{\mu}) = \Gamma_{\mu}
\Gamma_{D+1}, \: \:& \epsilon(\Gamma_{\mu}) = 0 \\
\Delta(\Gamma_{D+1}) = \Gamma_{D+1} \otimes \Gamma_{D+1},
\: \: \: & S(\Gamma_{D+1}) = \Gamma_{D+1}, \: \: &
\epsilon(\Gamma_{D+1}) = 1 \\
\end{array}
\end{equation}

The irreducible representations of $CH_q(D)$ are in one
to one correspondence with those of the Clifford algebra
$C(\eta)$ for all possible signatures of the metric $\eta$,
in D (D even) or
D+1 (D odd) dimensions respectively.
They are labelled by complex parameters
$\{\lambda_{\mu}\}_{\mu=1}^D$,
the eigenvalues of the Casimir operators $K_\mu=q^{E_\mu}$.
{}From
now on we restrict ourselves to the case $D$ even, $D=2M$.

The irreps of $CH_q(2M)$ are isomorphic to the tensor
product of $M$
$CH_q(2)$ irreps, being their dimension $2^M$. Thus,
a basis for $CH_q(2M)$ can be obtained from the
$CH_q(2)^{\otimes M}$ generators as follows
($\gamma_{\alpha}, E_{\alpha} (\alpha =1,2), \gamma_3$
$ \in CH_q(2)$):
\begin{eqnarray}
\Gamma_{2(n-1)+ \alpha} & = & \gamma_3 \otimes \cdots
\otimes \gamma_3
\otimes \stackrel{n)}{\gamma_{\alpha}} \otimes 1 \otimes
\cdots \otimes 1 \hspace{1cm} n=1,..,M; \; \alpha=1,2
\nonumber \\
E_{2(n-1)+ \alpha} & = & 1 \otimes \cdots \otimes 1
\otimes E_{\alpha} \otimes 1 \otimes \cdots \otimes 1 \\
\Gamma_{D+1} & = & \gamma_3 \otimes \cdots \otimes
\gamma_3 \nonumber
\end{eqnarray}
\noindent
The Hopf algebra $CH_q(2M)$ is related to the tensor
product
$CH_q(2)^{\otimes M}$ by a Drinfeld twist $B$ \cite{D}
\begin{equation}
\Delta_{\scriptscriptstyle CH_q(2M)} (g)= B
\Delta_{\scriptscriptstyle CH_q(2)^{\otimes M}} (a)
B^{-1} \hspace{6mm} \forall g \in CH_q(2M)
\end{equation}
where the operator
$B \in CH_q(2)^{\otimes M} \otimes CH_q(2)^{\otimes M}$
acting on the tensor product of two $CH_q(2M)$ irreps
is defined by
\begin{eqnarray}
& & B= (-1)^{F*F} \\
& & F*F= \sum_{1\leq j<i \leq M}
(1 \otimes \cdots \otimes \stackrel{i)}{f} \otimes
\cdots \otimes 1) \otimes (1 \otimes \cdots \otimes
\stackrel{j)}{f} \otimes \cdots \otimes 1) \nonumber
\end{eqnarray}
with $f=0$(boson),$1$(fermion) the fermion number for the
two vectors in a $CH_q(2)$ irrep.
The reason to introduce the operator $B$ in formula (6)
is that the comultiplication in $CH_q(2)^{\otimes M}$
treats each factor $CH_q(2)$ separatedly. This can be
represented by a twist between the $CH_q(2)$ pieces of a
$CH_q(2M)$ irrep.
Since one of the vectors in a $CH_q(2)$ irrep behaves as
a fermion, this twist has the effect of introducing some
signs that we represent by the operator $B$ (fig.1).

\begin{figure}[b]
\epsffile{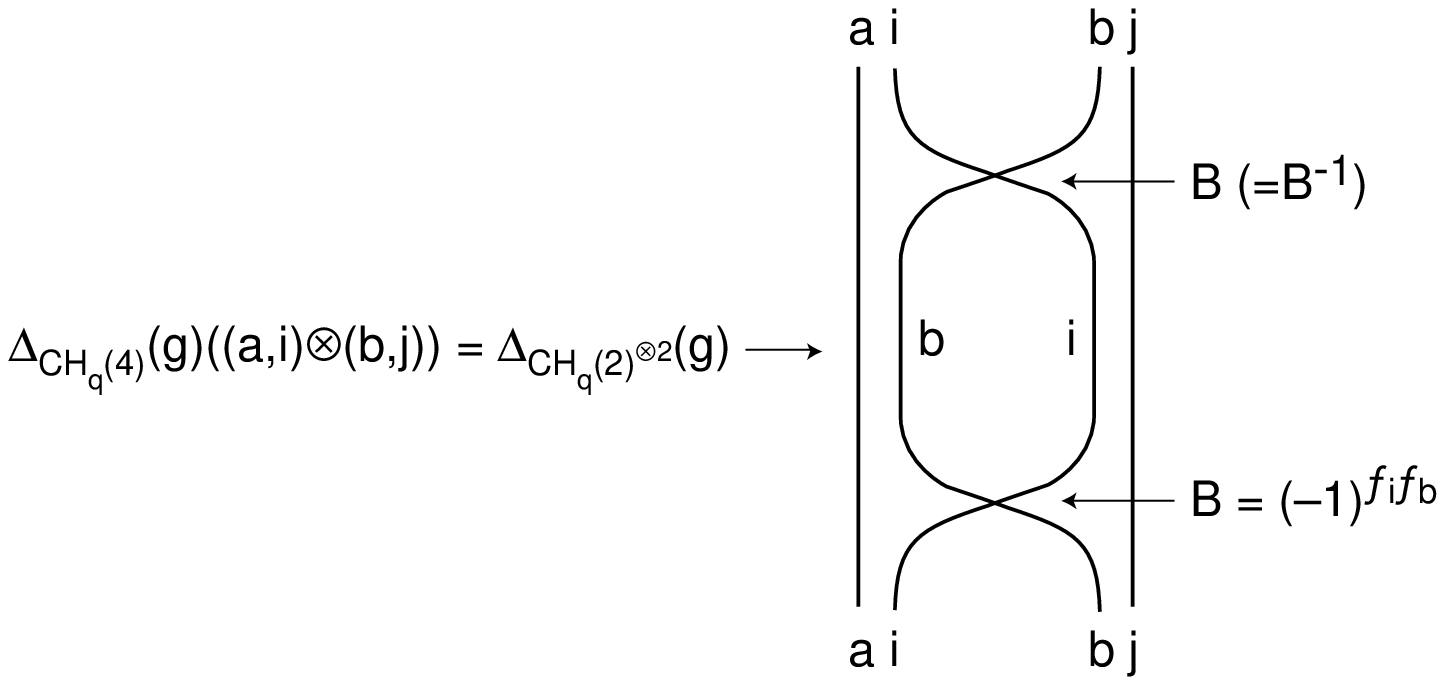}
\caption{Graphycal  representation  of  the  expression
(6) for $CH_q(4)$. $(a,i)$ denote the vectors in a
$CH_q(2)^{\otimes 2}$
irrep, the index $a$ corresponding to the first
$CH_q(2)$ and $i$ to the second.}
\end{figure}

Next we introduce a sort of affinization of the Hopf
algebra $CH_q(D)$.
The generators of this new algebra $\widehat{CH_q(D)}$
are $\Gamma_{\mu}^{(i)}$, $E_{\mu}^{(i)}$ ($i=0,1$) and
$\Gamma_{D+1}$ verifying (3) and (4) for each value of
$i$. We impose also that the anticommutator
$ \{ \Gamma_{\mu}^{(1)} , \Gamma_{\nu}^{(2)} \}$
belong to the center of
$\widehat{CH_q(D)}$ $ \forall \mu , \nu$.

Let's give now the explicit realization of
$\widehat{CH_q(2)}$. It is an useful example, and it
will provide us with the building blocks for any $D$.
A two-dimensional irrep $\pi_{\xi}$ of
$\widehat{CH_q(2)}$
is labelled by $\xi = (z,\lambda_1,\lambda_2) \in
C^3$ and reads as follows
\begin{eqnarray}
& \begin{array}{ccc}
\pi_{\xi}(\gamma^{(0)}_1) =
\left(\frac{\lambda_1^{-1}-\lambda_1}{q-q^{-1}}\right)^{1/2}
\left(
\begin{array}{cc} 0 & z^{-1} \\ z & 0 \\ \end{array} \right)
&, & \pi_{\xi}(\gamma^{(1)}_1) =
\left(\frac{\lambda_1-\lambda_1^{-1}}{q-q^{-1}}\right)^{1/2}
\left(
\begin{array}{cc} 0 & z \\ z^{-1} & 0 \\ \end{array} \right)
\\
& & \\
\pi_{\xi}(\gamma^{(0)}_2) =
\left(\frac{\lambda_2^{-1}-\lambda_2}{q-q^{-1}}\right)^{1/2}
\left(
\begin{array}{cc} 0 & -i z^{-1} \\ i z & 0 \\ \end{array}
\right) &, & \pi_{\xi}(\gamma^{(1)}_2) =
\left(\frac{\lambda_2-\lambda_2^{-1}}{q-q^{-1}} \right)^{1/2}
\left(
\begin{array}{cc} 0 & -i z \\i z^{-1} & 0 \\ \end{array}
\right) \\
\end{array} & \\ \nonumber \\
& \pi_{\xi}(\gamma_3) =
\left( \begin{array}{cc} 1 & 0 \\ 0 & -1 \\ \end{array}
\right)
\hspace{1cm} , \hspace{1cm}
\left. \begin{array}{lll} \pi_{\xi}(q^{E^{(0)}_1}) =
\lambda_1^{-1} & , &
\pi_{\xi}(q^{E^{(1)}_1}) = \lambda_1 \\
\pi_{\xi}(q^{E^{(0)}_2}) = \lambda_2^{-1} &,
& \pi_{\xi}(q^{E^{(1)}_2}) =
\lambda_2 \\ \end{array} \right.  \nonumber \\
\nonumber
\end{eqnarray}

For the affine $\widehat{CH_q(2M)}$ we can define a
straightforward
generalization of the expression (5). It allows to
introduce $M$
different affinization parameters $\{ z_n \}_{n=1}^{M}$,
one for each $\widehat{CH_q(2)}$ piece
\begin{eqnarray}
\Gamma_{2(n-1)+ \alpha}^{(i)} & = & \gamma_3 \otimes \cdots
\otimes \gamma_3 \otimes \gamma_{\alpha}^{(i)}
\otimes 1 \otimes \cdots \otimes 1 \hspace{1cm} n=1,..,M;
\; \alpha=1,2;  \; i=0,1 \nonumber \\
E_{2(n-1)+ \alpha}^{(i)} & = & 1 \otimes \cdots \otimes 1
\otimes
E_{\alpha}^{(i)} \otimes 1 \otimes \cdots \otimes 1 \\
\Gamma_{D+1} & = & \gamma_3 \otimes \cdots \otimes
\gamma_3 \nonumber \\
\nonumber
\end{eqnarray}

The intertwiner $R-$matrix for two irreps with labels
$\xi = \{ z_{n},\lambda_{2n-1},\lambda_{2n} \}_{n=1}^{M}$
is defined by the
condition
\begin{equation}
R_{\xi_1 \xi_2} \Delta_{\xi_1 \xi_2}(g) = \Delta_{\xi_2
\xi_1}(g) R_{\xi_1 \xi_2} \;\;\; \forall g \in
\widehat{CH_q(2M)}
\end{equation}
\noindent
with $\Delta_{\xi_1 \xi_2}= \pi_{\xi_1} \otimes \pi_{\xi_2}
(\Delta)$. Since (6) remains true for any element
$g \in \widehat{CH_q(2M)}$, the intertwiner $R-$matrix
between two irreps (which furthermore satisfies the
Yang-Baxter equation) is given by  \cite{D}
\begin{equation}
R_{\scriptscriptstyle CH_q(2M)}(u_{1},..,u_{M}) = B \:
R_{\scriptscriptstyle CH_{q}(2)^{\otimes M}}
(u_{1},..,u_{M}) \: B^{-1} \\
\end{equation}
\[  R_{\scriptscriptstyle CH_{q}(2)^{\otimes M}}
(u_{1},..,u_{M})
=R_{\scriptscriptstyle CH_q(2)}^{(1)}(u_{1}) \ldots
R_{\scriptscriptstyle CH_q(2)}^{(M)}(u_{M}) \]
The matrices $R_{\scriptscriptstyle CH_q(2)}^{(n)} =
R_{\xi_{1}^{(n)}
\xi_{2}^{(n)}}$ ($\xi^{(n)}=(z_{n},\lambda_{2n-1},
\lambda_{2n})$) are the $\widehat{CH_q(2)}$ intertwiners
\begin{eqnarray}
& & \begin{array}{ll}
R_{00}^{00}=1-e(u_{n})e_{1}e_{2} \; , & R_{11}^{11}=
e(u_{n})-e_{1}e_{2} \\
R_{01}^{10} = e_{1} - e(u_{n})e_{2} \; , & R_{10}^{01}=
e_{2} - e(u_{n})e_{1} \\
\end{array} \\
& & R_{01}^{01}  = R_{10}^{10}=(e_{1}sn_{1})^{1/2}
(e_{2}sn_{2})^{1/2}(1-e(u_{n}))/sn(u_{n}/2)  \nonumber \\
& & R_{00}^{11}  = R_{11}^{00}=-ik(e_{1}sn_{1})^{1/2}
(e_{2}sn_{2})^{1/2}(1+e(u_{n}))sn(u_{n}/2) \nonumber
\end{eqnarray}
where $e(u_n) =  cn(u_n) +  i sn(u_n)$ is the elliptic
exponential of modulus $k_n$,
$e_{i}=e(\psi_{i}^{n}), sn_{i}=sn(\psi_{i}^{n})$
($i=1,2$) and $u_n, \psi_{i}^{n}$ are
elliptic angles depending on the labels $\xi_{i}^{(n)}$
(see ref.\cite{U} for details).

There is a
constraint on the irrep labels so that (12) be indeed
their intertwiner
\begin{equation}
\frac{2 (\lambda_{2n-1} - \lambda_{2n})}
{(1-\lambda_{2n-1}^2)^{1/2}
(1-\lambda_{2n}^2)^{1/2}(z_{n}^2 - z_{n}^{-2})} = k_{n} \; ,
\hspace{5mm} n=1,..,M
\end{equation}

All the $R_{CH_q(2)}^{(n)}$ matrices are independent and
commute among them. It's remarkable that the spectral
curve (13) of irreps that
admit an intertwiner is parametrized by $M$ independent
elliptic
moduli $k_n$. Indeed, some of them can be in the elliptic
regime and others in the trigonometric ($k \! = \! 0$).
The matrix $R_{CH_q(2M)}$ can be thought of
as the scattering matrix for objects composed of M
different kinds of particles. There is real interaction
when two equal particles scatter from each other, given
by $R_{CH_{q}(2)}^{(n)}$; otherwise
there is only a sign coming from their statistics and
represented by the operator $B$ (fig 2).

Finally, note that the $R-$matrix (12) coincides with the
Boltzmann weights for the most
general 8-vertex free fermionic solution to the Yang-Baxter
equation in non zero magnetic field \cite{F,BS}.

\begin{figure}
\epsffile{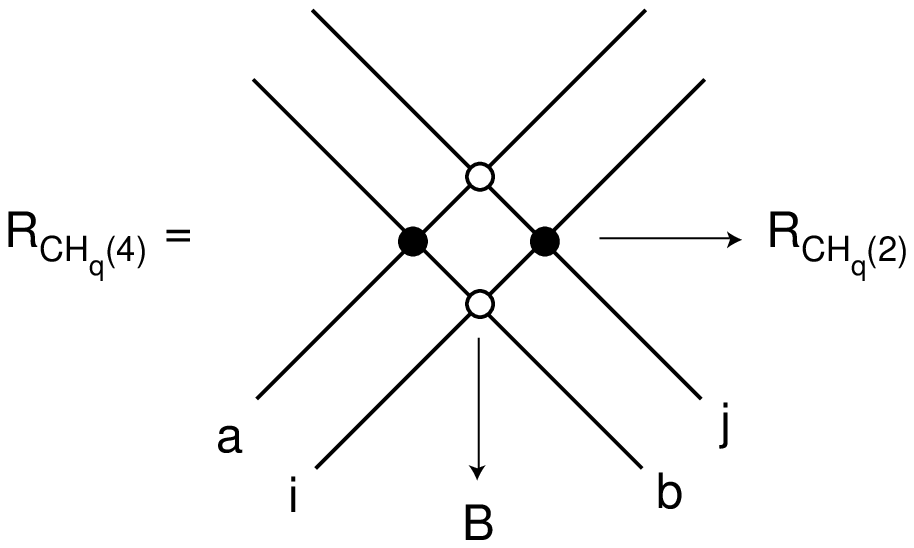}
\caption{Graphycal representation of the $CH_q(4)$
$R-$matrix.}
\end{figure}

\subsection*{3. Extended supersymmetry}
\indent

In order to analyze the connection of $\widehat{CH_q(2M)}$
with supersymmetry algebras,
we will study the limit in which the $R-$matrix (12)
becomes trigonometric.
Let us consider first the case $D=2$ in detail. This case
turns out to be related to an $N=2$ (2 supersymmetry charges)
integrable Ginzburg-Landau model.
We shall also give an heuristic motivation for the
construction of the Hopf algebra $\widehat{CH_q(2)}$ based
on its trigonometric 6-vertex limit.

The 6-vertex free fermionic solutions are given by the
intertwiner $R-$matrix between nilpotent irreps of
$U_{\epsilon}(\widehat{sl(2)})$, $\epsilon^4 \! = \! 1$
($\Rightarrow \epsilon \! = \! i$) \cite{N}.
In a $U_{\epsilon=i}(sl(2))$ nilpotent irrep the values of
the special Casimirs are $Q_{\pm}^2 \! = \! 0$ ($Q_{\pm}=
S_{\pm} \epsilon^{\pm H/2}$) and $K^2 \! = \! \lambda^2$
arbitrary ($K\!=\! \epsilon^{H}$); namely, they are the
highest weight case of the cyclic irreps.
Furthermore when $\epsilon^4=1$ the anticommutator
$\{ Q_{+},Q_{-} \}$ also belongs to the center, suggesting
the connection
with a Clifford algebra through the mixing of the positive
and negative root generators $Q_{\pm}$. The total fermion
number is conserved in the 6-vertex solutions to the
Yang-Baxter equation, but it is not in the elliptic regime.
Hence a non trivial mixing is needed to represent the
elliptic regime. The Hopf algebra $CH_q(2)$ assigns
different {\it central elements}
$\left[ E_{1} \right]_q , \left[ E_{2} \right]_q$
to the square of the generators $\gamma_{1}$, $\gamma_{2}$
respectively, in such a way that the mixing can only be
undone
(trigonometric limit) when $E_{1} \! = \! E_{2} \! = \!E$.
It implies $k=0$ in (13). For the affine $\widehat{CH_q(2)}$
this limit leads to
$U_{\epsilon=i}(\widehat{sl(2)})$ (this statement is only
rigurous for the affine case): i.e. $R_{CH_q(2)}$
becomes the $R-$matrix intertwiner for
$U_{\epsilon=i}(\widehat{sl(2)})$, provide the labels of
the two algebras are related by $\lambda \! = \! q^E$.

Using the generators
$Q_{\pm},\overline{Q}_{\pm} \in U_{\epsilon =i}
(\widehat{sl(2)})$, we
can define an $N=2$ supersymmetry algebra with topological
extension $T_{\pm}$ \cite{BL,V}
\begin{eqnarray}
& & Q_{\pm}^{2} = \overline{Q}_{\pm}^{2} =
\{ Q_{\pm} ,\overline{Q}_{\pm} \} =0  \nonumber \\
& & \{ Q_{\pm} , \overline{Q}_{\mp} \} =2T_{\pm} \; ,
\hspace{5mm}
\mid \! T_{\pm} \! \mid = \left[ E \right]_q \\
& & \{ Q_{+} , Q_{-} \} =2m \: z^{2} \; , \hspace{5mm}
\{ \overline{Q}_{+} ,
\overline{Q}_{-} \} =2m \: z^{-2} \nonumber
\end{eqnarray}
satisfying the Bogomolnyi bound $\mid \! T_{\pm} \! \mid =m$.
The free fermionic condition (1) ensures the $N=2$ invariance
of the $R-$matrix.
Moreover, the $N=2$ part of the scattering matrix for the
solitons of the Ginzburg-Landau superpotential
$W=X^{n+1} /(n+1)- \beta X$ \cite{FI} is given by $R-$matrices
of $U_{\hat{q}}(\widehat{gl(1,1)})$ with $\hat{q}^{2n}\!=\!1$
\cite{SR}, or equivalently by those of $U_{\epsilon = i}
(\widehat{sl(2)})$ between nilpotent irreps with labels
$\lambda\!=\! \hat{q}$ \cite{M}.

The Ginzburg-Landau models have a particular importance in
the context of $N=2$ supersymmetry, since they allow to
classify a wide variety of $N=2$ superconformal field
theories \cite{VW}.
Of great interest are the relevant perturbations of these
theories giving massive integrable models, as happens for
the superpotential
$W(X)=X^{n+1} /(n+1)- \beta X$.
We would like now to make plausible in this context why the
supersymmetry algebra (14) has a non-trivial comultiplication.
In a $N=2$ Ginzburg-Landau model, the superpotential enters
explicitly in the SUSY conmutators through
\begin{eqnarray}
& & \{ Q_{+}, \overline{Q}_{+} \} = \Delta W  , \hspace{1cm}
\{ Q_{-}, \overline{Q}_{-} \} = \Delta W^{*} \\
& & \Delta W = W(X^{j})-W(X^{i}) \nonumber
\end{eqnarray}
with $X(-\infty) \! = \! X^i$, $X(\infty) \! = \! X^j$ and
$X^i, X^j$ minima of $W$. Let's call $K_{(i,i+l)}$ the
soliton going from $X^i$ to $X^j$, where $l=j-i$.
It is straighforward to see that $\Delta W$ depends on
both $l$ and $i$. Naively, the dependence in $i$ was
not expected since
all the solitons with the same $l$ are equivalent. For the
superpotential proposed it is possible to obtain a
supersymmetric algebra without this dependence, at the price
of reabsorbing it in a non-trivial quantum group
comultiplication
\begin{eqnarray}
\Delta (Q_{\pm}) = q^{\pm E} \gamma_{3} \otimes
Q_{\pm} + Q_{\pm} \otimes {\bf 1} & &  \\
\Delta (\overline{Q}_{\pm}) = q^{\mp E} \gamma_{3} \otimes
\overline{Q}_{\pm} + \overline{Q}_{\pm}
\otimes {\bf 1} & & \nonumber
\end{eqnarray}
\indent
On the other hand, it is worth noting the relation of (16)
with the fermion number of the solitons.
In the solitonic sectors, the fermion number operator
acquieres a fractional constant piece due to the interaction
of the fermionic degrees of freedom with the solitonic
background. The fractional piece of the fermion number in
a soliton sector $K_{(i,j)}$, is given by \cite{FI2,G}
\begin{equation}
f=- \frac{1}{2\pi} ( \: Im \: \ln{ \: W^{''} (X)} \:)
\mid_{X^i}^{X^j} \: \:
= \: \: \frac{s}{n}  \; \hspace{1cm} s=1,...,n-1
\end{equation}
The relation with $CH_q(2)$ labels is
$q^{E}\!=\!e^{i \pi s/n}$. Therefore $q^{\pm E} \! \gamma_3$
in (16) would be the analog of $e^{\pm i \pi F}$, with $F$
the fermion number operator. This interpretation fails for
$\Delta(\overline{Q}_{\pm})$, where the signs are
interchanged, leading in fact to a quantum group structure
instead to a Lie superalgebra.

Let us return to buiding extended supersymmetry algebras
from the general $\widehat{CH_q(2M)}$, in the same sense
as above.
The trigonometric limit of
$\widehat{CH_q(2M)}$ is obtained as an independent
trigonometric limit
in each $\widehat{CH_q(2)}$ piece. Then the affine Hopf
algebra
$\widehat{CH_q(2M)}$ becomes
in essence the anticommuting tensor product of $M$
$U_{\epsilon=i}(\widehat{sl(2)})$ factors, each with
its own spectral parameter.
Imposing that the eigenvalues of all the central
charges $E_i$ and the spectral parameters $z_i$
($i=1,..,M$) coincide,
we get $M$ copies of the same structure (14),
$\{ Q_{\pm}^{(i)}, \,
\overline{Q}_{\pm}^{(i)}, \, T_{\pm}^{(i)} \! =
\! T_{\pm} \}_{i=1}^{M}$.
Therefore we find an $N=2M$ supersymmetry algebra with
$M$ topological charges. Indeed, the dimension of a
$\widehat{CH_q(2)}$ irrep is $2^M$ as is needed to
saturate the Bogomolnyi bound $\mid \! T_{\pm}^{(i)} \!
\mid=\mid \! T_{\pm} \! \mid=m$.

Besides, we have seen that the $\widehat{CH_q(2M)}$
irreps can be thougth of as collections of M
independent solitons $\widehat{CH_q(2)}$.
Let us consider the more general trigonometric limit
with equal values of the central charges $E_i$, but
arbitrary spectral parameters $z_i$ ($i=1,..,M$).
Then the charges
\begin{equation}
Q_{\pm}^{T}=\sum_{i=1}^{M} Q_{\pm}^{(i)} \hspace{6mm},
\hspace{6mm}
\overline{Q}_{\pm}^{T} = \sum_{i=1}^{M}
\overline{Q}_{\pm}^{(i)}
\end{equation}
verify the commutation relations of $N=2$ supersymmetry
(14). In fact, (14) is satisfied even if we allow
different central charges $E_i$ . However, in this case
the comultiplication doesn't
preserve the expression (18) of
$Q_{\pm}^{T},\overline{Q}_{\pm}^{T}$.

\subsection*{4. Generalized $XY$ spin chains}
\indent

The quantum group structure plays an important role
in 2-dimensional statistical models, since $R-$matrix
intertwiners provide systematic solutions to the
integrability condition, the Yang-Baxter equation. In
this way integrable models can be built associated to
a quantum group, allowing to connect integrability with
an underlying symmetry principle. As noted above, the
intertwiner $R-$matrix for the Clifford-Hopf algebra
$\widehat{CH_q(2)}$ reproduces the 8-vertex free fermion
model in magnetic field. In this section we will analyze
the model defined by the algebras $\widehat{CH_q(D)}$
for general $D=2M$. Following the transfer matrix method,
the study of a 2-dimensional statistical model is
equivalent to that of its corresponding spin chain.
The L-site
hamiltonian for a periodic chain defined by the
$\widehat{CH_q(2M)}$ Hopf algebras is given by (provided
that $R(0)\!=\! {\bf 1}$)
\begin{eqnarray}
& & H= \sum_{j=1}^{L} i \frac{\partial}{\partial u}
R_{j,j+1}(u) \mid_{u=0} \\
& & H= \sum_{j=1}^{L} \sum_{n=1}^{M} \{ (J_{x}^{n}
\sigma_{x,j}^{n}
\sigma_{x,j+1}^{n} + J_{y}^{n} \sigma_{y,j}^{n}
\sigma_{y,j+1}^{n})  \sigma_{z,j}^{n+1} ... \sigma_{z,j}^{M}
\sigma_{z,j+1}^{1} ... \sigma_{z,j+1}^{n-1} \: + \: h^{n}
\sigma_{z,j}^{n} \}  \nonumber \\
\nonumber
\end{eqnarray}
where $\sigma_{a}^{n}$ ($a=x,y,z \, , \; n=1,..,M$) are
$M$ sets of Pauli matrices, and the constants
$J_{x}^{n},J_{y}^{n},h^{n}$ depend on the quantum labels
of the irreps whose intertwiner is $R$
\begin{eqnarray}
J_{x}^{n} & = & 1+ \Gamma^{n} \; , \; J_{y}^{n} \, = \, 1-
\Gamma^{n} \hspace{1cm} n=1,..,M \nonumber \\
\Gamma^{n} & = &  k_{n}sn(\psi^{n}) \\
h^{n} & = & 2 cn(\psi^{n}) \nonumber
\end{eqnarray}
The requirement $R(0)= {\bf 1}$
implies $\psi_{1}^{n}=\psi_{2}^{n}=\psi^{n}$.

The hamiltonian (19) can be diagonalized through a
Jordan-Wigner
transformation and its excitations behave as free fermions
(massless when
$J_{x}^{n}=J_{y}^{n}$ massive otherwise).
This model provides $M$ groups of Pauli matrices
$\sigma_{a,j}^{n}$ ($a=x,y,z$)
for each site $j$ on the chain, so it behaves as having
$M$ layers
with an $XY$ model defined in each layer.
The factors $(\sigma_{z,j}^{k+1}
... \sigma_{z,j}^{M} \sigma_{z,j+1}^{1} ...
\sigma_{z,j+1}^{k-1})$
make the fermionic excitations on different layers
anticonmute.
Thus the algebra $\widehat{CH_q(2M)}$ provides a way
to put different
non-interacting fermions in a chain with a quantum group
interpretation.

When $M \! = \! 1$, $H$
reduces to the hamiltonian of an $XY$ Heisenberg chain in
an external
magnetic field $h$, that is the spin chain asociated with
the 8-vertex free fermion model \cite{LSM}
\begin{equation}
H= \sum_{j=1}^{L} \{ J_{x} \sigma_{x,j}
\sigma_{x,j+1} + J_{y} \sigma_{y,j}
\sigma_{y,j+1}  + h \sigma_{z,j} \}  \\
\nonumber
\end{equation}

The aim of this section is to show that the model above
is equivalent under some restrictions to the generalized
integrable $XY$
chain proposed and solved in ref.\cite{S},
\begin{equation}
\widetilde{H} = - \sum_{k=1}^{K}
\sum_{j=1}^{L'}  (\tilde{J}_{x}^{k}
\sigma_{x,j} \sigma_{x,j+k} + \tilde{J}_{y}^{k} \sigma_{y,j}
\sigma_{y,j+k} ) \sigma_{z,j+1} \ldots \sigma_{z,j+k-1} \,
+ \, h \sum_{j=1}^{L'}  \sigma_{z,j}   \\
\nonumber
\end{equation}
finding in this way a quantum group structure for this
integrable model.
The hamiltonian (22) can also be diagonalized with a
Jordan-Wigner transformation and its quasi-particles behaves
as free fermions.
The main application of the generalized $XY$ model is the
problem of covering a surface with horizontal and vertical
dimers. Indeed, the ground state of $\widetilde{H}$ for a
particular choice of parameters reproduces the two-dimensional
pure dimer problem \cite{S}, first solved in terms of a
Pfaffian \cite{K}.

To see the relation between $H$ and $\widetilde{H}$, let us
choose identical $XY$ models on each layer of the former
chain
\begin{equation}
J_{x}^{n}=J_{x} \: , \hspace{8mm} J_{y}^{n}=J_{y} \: ,
\hspace{8mm} h^{n}=h \hspace{8mm}  n=1,..,M
\end{equation}
and rearrange the spin labels to form a single-layer chain
\begin{equation}
\sigma_{a,j}^{n}=\sigma_{a,j+n}   \hspace{12mm} n=1,..,M \,
; \; a=x,y,z
\end{equation}
Then the hamiltonians $H$ and $\widetilde{H}$ coincide if
we set in the latter
\begin{equation}
\tilde{J}_{x}^{k} = - J_{x} \delta_{M,k} \, , \hspace{8mm}
\tilde{J}_{y}^{k} = - J_{y} \delta_{M,k}
\hspace{1cm} k=1,..,K
\end{equation}
\indent
The general $\widetilde{H}$ (22) is obtained by adding
hamiltonians $H^{(M)}$ derived from $\widehat{CH_q(2M)}$
$R-$matrices.
The fact that this sum is also solvable
relies on setting equal parameters
in each $H^{(M)}$ (this is the same condition
that leads to $N=2M$ supersymmetry in the
trigonometric limit of $\widehat{CH_q(2M)}$).
Therefore, the affine quantum Clifford Hopf algebras
$\widehat{CH_q(2M)}$ encode the hidden quantum group
for the generalized $XY$ spin chain (22).

\subsection*{5. Comments}
\indent

We have studied the quantum Clifford algebras
$\widehat{CH_q(2M)}$ in connection with extended
supersymmetry and with statistical integrable models.

It is worth noting that the hamiltonian derived from
$\widehat{CH_q(4)}$
in the trigonometric regime and without magnetic field, is
the limiting case $U \rightarrow \infty$ of the two layer
chain \cite{B}:
\begin{equation}
H= - \frac{1}{2} \sum_{j=1}^{L} \{ ( \sigma_{x}^{j}
\sigma_{x}^{j+1} + \sigma_{y}^{j} \sigma_{y}^{j+1} )
(1-U \tau_{z}^{j+1}) + ( \tau_{x}^{j} \tau_{x}^{j+1} +
\tau_{y}^{j} \tau_{y}^{j+1} ) (1- U \sigma_{z}^{j}) \}
\end{equation}
The coupling between the two layers in this model implies real
interaction, so the excitations are not free fermions, and the
ground state presents spontaneous magnetization
(if $U \neq 0, \infty$).
It still can be solved by Bethe Ansatz techniques, but a
$R-$matrix interpretation
for it is not known. The algebra $\widehat{CH_q(4)}$ gives us
a simple way of coupling two $XY$ models. Perhaps it would be
possible
to twist (may be in a way related to a quantum deformation
proposed recently for the Clifford algebras \cite{C}) and
break the full set of generators to a
shorter set giving a quantum group structure for this model.

We have built extended supersymmetric algebras
from the $\widehat{CH_q(2M)}$ generators in the trigonometric
limit.
The Clifford Hopf algebras can be thought of as elliptic
generalizations of supersymmetry (the anticommutators of
charges that
give the momentum $P$ and $\overline{P}$ get deformed in
the
elliptic case, but are still central elements). It would be
interesting to
analize what deformation of the Poincar\'e group one gets
in such a way.

\vspace{1cm}

{\bf Acknowledgements}

\indent

The author would like to thank A. Berkovich, R. Cuerno,
C. G\'omez
and G. Sierra for discussions and encouragement. This work is
supported by M.E.C. fellowship AP91 34090983.

\end{document}